# Modeling Solar Spectral Irradiance (SSI) from Iron lines using the COronal DEnsity and Temperature (CODET) model version 1.1

Jenny M. Rodríguez-Gómez [1]

[1]*The Catholic University of America located at NASA Goddard Space Flight Center, Greenbelt, MD 20771*

ABSTRACT

The COronal DEnsity and Temperature (CODET) model is a physics-based model (Rodríguez-Gómez 2017; Rodríguez Gómez et al. 2018). This model uses the relationship between the magnetic field, density, temperature, and EUV emission. This model provides mean daily Solar Spectral Irradiance time series in EUV wavelengths in long time scales from days to solar cycles. The current manuscript presents the updated/new CODET model version 1.1. It uses observational datasets from SDO/EVE MEGS-A at 28.4 nm and 21.1 nm wavelengths to obtain the goodness-of-fit between them and the modeled SSI from April 30, 2010, to May 26, 2014. The model described well observational data during that period, with less than 20% error in both wavelengths. Additionally, SSI predictions are provided using the new model parameters from July 1996 to October 2024, where SOHO/MDI and SDO/HMI photospheric magnetic field data are available. These predictions were compared with GOES/EUVS at 28.4 nm and mean daily values of SDO/AIA at 21.1 nm data, with errors of ∼ 26% and 42% for 28.4 nm and 21.1 nm, respectively. The error analysis for model fitting and predictions shows how accurate the model predictions are. Thus, the CODET model provides a reliable estimate of the Solar Spectral Irradiance time series in EUV wavelengths where no observational data is available, e.g., after the SDO/EVE MEGS-A era.

## 1. INTRODUCTION

Solar Spectral Irradiance (SSI) provides information about the physical, chemical, thermal, and dynamical properties of the solar atmosphere. Most of the Extreme UltraViolet (EUV) emerges from the solar corona and transition region, e.g., Iron lines like Fe XV 28.4 nm and Fe XIV 21.1 nm describe emission from the solar corona. Solar emission in EUV wavelengths is the primary source of energy input to planetary atmospheres and plays an indispensable role in the atmospheric structure and composition. In particular, EUV emission is responsible for the photochemistry process in the planetary upper atmosphere (de Oliveira et al. 2024; Fox et al. 2008). It has an essential role in the Earth's atmosphere, driving the dynamics in the Ionosphere - Thermosphere -Mesosphere (ITM) system. SSI variability is an important driver in the heating of the terrestrial upper atmosphere (Ermolli et al. 2013; Reiss et al. 2023), especially the thermosphere (Lukianova & Mursula 2011). In addition, the Ionosphere shows a strong correlation with solar EUV variability (Cander 2019; Anderson & Hawkins 2016). An example is the Travel Ionospheric Disturbances (TIDs) or scintillations related to the radio wave propagation affecting GPS signals (Kintner et al. 2007; Hern´andez-Pajares et al. 2012). Additionally, SSI at 28.4 nm plays an important role in the ionospheric response even if no flares or strong emissions are involved (Bekker et al. 2024), providing an excellent proxy to study the ionosphere.

Solar Spectral Irradiance below 300 nm does not reach the Earth's surface because this irradiance is attenuated through scattering and absorption by the Earth's atmosphere (Floyd et al. 2005). Due to these effects, accurate SSI measurements must be made from space (Thuillier et al. 2022). However, observational challenges like instrumental degradation, cross-calibrations, overlapping observations between the different instruments (Carlesso et al. 2022), and observational gaps between others make it difficult.

The EUV Variability Experiment (EVE) onboard Solar Dynamic Observatory (SDO) (Hock et al. 2012; Woods et al. 2012; Chamberlin et al. 2007) has been providing EUV data since 2010. EVE has two detectors, MEGS-A and MEGSB covering 6–33.3 nm and 33.3–105 nm wavelength intervals, respectively. However, a malfunction in 2014 affected the EVE/MEGS-A detector. Thus, EVE/MEGS-A data only cover the interval from April 30, 2010, to May 26, 2014.

Corresponding author: Jenny Marcela Rodríguez-Gómez rodriguezgomez@cua.edu, jenny.m.rodriguezgomez@nasa.gov

Nevertheless, long-term and accurate measurements of solar irradiance are needed to understand its effect on different time scales on the Ionosphere - Thermosphere -Mesosphere (ITM) Earth system. Thus, SSI irradiance models are needed to understand their variations over long time scales and where observational gaps exist.

In general, SSI models come from theoretical and empirical descriptions, e.g., estimates of the temperature structure of the solar atmosphere, radiative transfer, non-local thermodynamic equilibrium, and magneto-hydrodynamics to simulate magnetized plasma (Erickson et al. 2021), proxy models using sunspot numbers (Krivova & Solanki 2008), and using the photospheric magnetic field. Since the solar magnetic field drives the solar irradiance variability, it is possible to use the magnetic field to model the Solar Spectral Irradiance (SSI) (Reiss et al. 2023). For example, the CODET model provides SSI in EUV wavelengths (Rodríguez-Gómez 2017; Rodríguez Gómez et al. 2018, 2019), and F10.7, bands of FUV irradiance and solar indices like Mg II core-to-win ratio (Warren et al. 2021; Henney et al. 2015, 2012).

This work presents the modeling of Solar Spectral Irradiance (SSI) from Iron lines 28.4 nm and 21.1 nm using the COronal DEnsity and Temperature (CODET) model version 1.1. In particular, the updated version of the CODET model v1.1 is described in Section 2. The main results are presented in Section 3. Finally, Section 4 contains a discussion and concluding remarks.

## 2. **THE CORONAL DENSITY AND TEMPERATURE (CODET) MODEL VERSION 1.1**

The COronal DEnsity and Temperature (CODET) model version 1.0 was developed by Rodríguez-Gómez (2017). It is a physics-based model that provides coronal density and temperature estimations, emission maps, and daily Solar Spectral Irradiance (SSI) time series in EUV wavelengths (19.3 nm and 21.1 nm). This model retrieves and allows the analysis of the SSI irradiance variability in long time scales from days to solar cycles, primarily through the mean fulldisc intensity. The underlying assumption is that a relationship exists between magnetic field, density, temperature, and EUV emission (Rodríguez Gómez et al. 2018).

The CODET model version 1.0 and 1.1 use the Surface Flux Transport model of Schrijver (2001). It incorporates observations and statistical properties of the magnetic field structures observed on the solar surface using full-disc magnetogram data from MDI/SOHO (Scherrer et al. 1995) and HMI/SDO (Scherrer et al. 2012). These data are used as boundary conditions for a series of Potential Field Source Surface (PFSS) extrapolations (Schrijver 2001; Schrijver & De Rosa 2003).

The density and temperature distribution are described as a function of the magnetic field using scaling laws described by Rodr´ıguez G´omez et al. (2018). These plasma density and temperature profiles are subsequently used as input for the emission model to retrieve the Solar Spectral Irradiance in EUV. The emission model is based on the emission measure formalism, assuming that the emission lines are optically thin. The intensity at a specific wavelength can be described mainly by the contribution function $G(\lambda,T)$ from the CHIANTI atomic database and the density $N^2$. CODET model version 1.0 used CHIANTI atomic database 8.0, ionization, and abundance models from Mazzotta et al. (1998) and Meyer (1985). In contrast, the updated model version uses CHIANTI atomic database [1] 10.0.2 and ionization and abundance models from Scott et al. (2015b,a). The SSI at a specific wavelength is the full-disk mean intensity measure at a specific heliocentric distance, e.g., 1 AU. The CODET version 1.0 intensity maps at each height from PFSS extrapolation. While CODET version 1.1 provides full-disc intensity maps. They are obtained as the sum of the modeled intensity at each layer from 1.0 $R_\odot$ to 2.5 $R_\odot$ and integrated line of sight at 1 AU.

Both versions of the CODET model use an optimization algorithm Pikaia[2] to search for the best-fit model parameters. A brief description of the CODET model like density and temperature scaling laws and parameters, is presented in the appendix (more details on Rodríguez Gómez et al. (2018)). This algorithm allows comparing observed and modeled irradiance through goodness-of-fit ($\chi^2$). Although the methodology from the original model is similar, a significant change was included in this version. Specifically, the SSI used to find the goodness-of-fit is real observational data from SDO/EVE and the inclusion of the 28.4 nm wavelength. The CODET model version v.1.0 (Rodríguez Gómez et al. 2018) uses TIMED/SEE (Woods et al. 2008; Hock & Eparvier 2008; Woods et al. 2005) data to compare model outputs in 19.5 nm and 21.5 nm. However, SSI from TIMED/SEE differs from observational data, e.g., the EUV Variability Experiment (EVE) onboard SDO (Hock et al. 2012; Woods et al. 2012; Chamberlin et al. 2007), because TIMED wavelengths shorter than 27 nm, 115–120 nm, and 122–129 nm are populated using models driven by SEE measurements[3]. Thus, comparing SSI at 21.5 nm and 28.5 nm from TIMED/SEE with SDO/EVE at 21.1 nm and 28.4 nm from April 30, 2010 to May 26, 2014. The mean intensity ratio from $\frac{EVE_{28.4}}{TIMED_{28.5}} \sim 9$ and $\frac{EVE_{21.1}}{TIMED_{21.5}} \sim 3$ shows an enormous difference between SDO/EVE and TIMED/SEE. Thus, to improve the model outputs and provide reliable SSI time series, the CODET version 1.1 uses observational data from EVE/SDO level 3 version 7.0 at 28.4 nm and 21.1 nm. Figure 1 shows the schematic description of the COronal DEnsity and Temperature (CODET) model version 1.1.    The observational datasets are shown in green rectangles.

---

[1] https://www.chiantidatabase.org/
[2] http://virtualrat.org/software/beluga

Orange rectangles correspond to models, algorithms and tests, which are depicted as yellow rectangles. The CODET model v1.1 routines and outputs are described with blue rectangles.

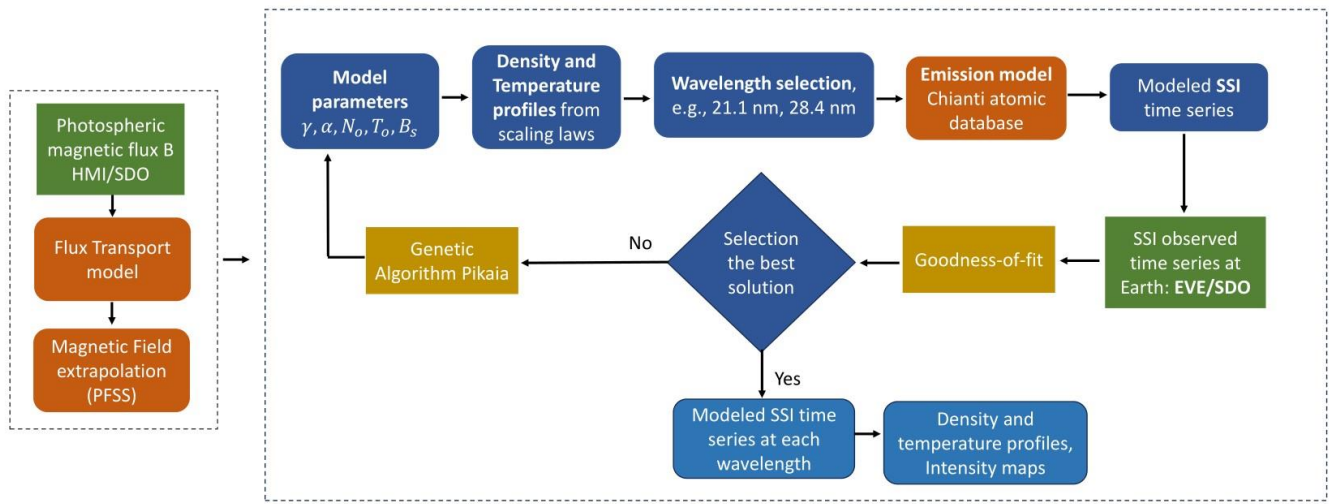

**Figure 1.** Schematic description of the COronal DEnsity and Temperature (CODET) model version 1.1.

The best model selection consists of different features, as was presented before by Rodríguez Gómez et al. (2018), e.g., The goodness-of-fit is determined by the $\chi^2$ value obtained from the comparison between observational and modeled SSI (more details in Appendix A) and by checking whether the density and temperature mean profiles through the solar atmosphere shows the expected behavior and value limits.

The CODET model version 1.1 focused on modeling the Solar Spectral Irradiance (SSI) in two specific wavelengths, Fe XV 28.4 nm and Fe XIV 21.1 nm during the EVE/SDO MEGS-A period observation, and the SSI prediction is presented from July 01, 1996, to October 28, 2024.

## 3. RESULTS

This section presents the main results. That is the Solar Spectral Irradiance (SSI) modeling from April 30, 2010, to May 26, 2014, SDO/EVE MEGS-A observation period, full-disc intensity, and density maps. In addition, Solar Spectral Irradiance CODET model v1.1 predictions from July 01, 1996, to October 28, 2024. Also, an error analysis using the Mean Absolute Percentage Error and the relative error is presented, as well as a comparison between model versions. Different wavelength combinations were explored to fit simultaneously in the model. However, Fe XV 28.4 nm and Fe XIV 21.1 nm wavelengths show better results due to their similar temperature formation, $\log_{10}$ T = 6.30 K and $\log_{10}$ T = 6.27 K, respectively. In this case, a day every three days were selected to find the goodness-of-fit during April 30, 2010, to May 26, 2014, SDO/EVE MEGS-A period. Figure 2 shows Solar Spectral Irradiance modeling using the CODET model version 1.1 at 28.4 nm and 21.1 nm from April 30, 2010, to May 26, 2014, and the observed irradiance from EVE/SDO. Table 1 lists the model parameters obtained from the simultaneous fitting of SSI in 28.4 nm and 21.1 nm wavelengths. As well as the Pikaia algorithm specifications such as the population size and generations to find the best solution and the goodness-of-fit ($\chi^2$) following the description presented in Figure 1.

---

[3] https://lasp.colorado.edu/lisird/data/timed_see_ssi_l3

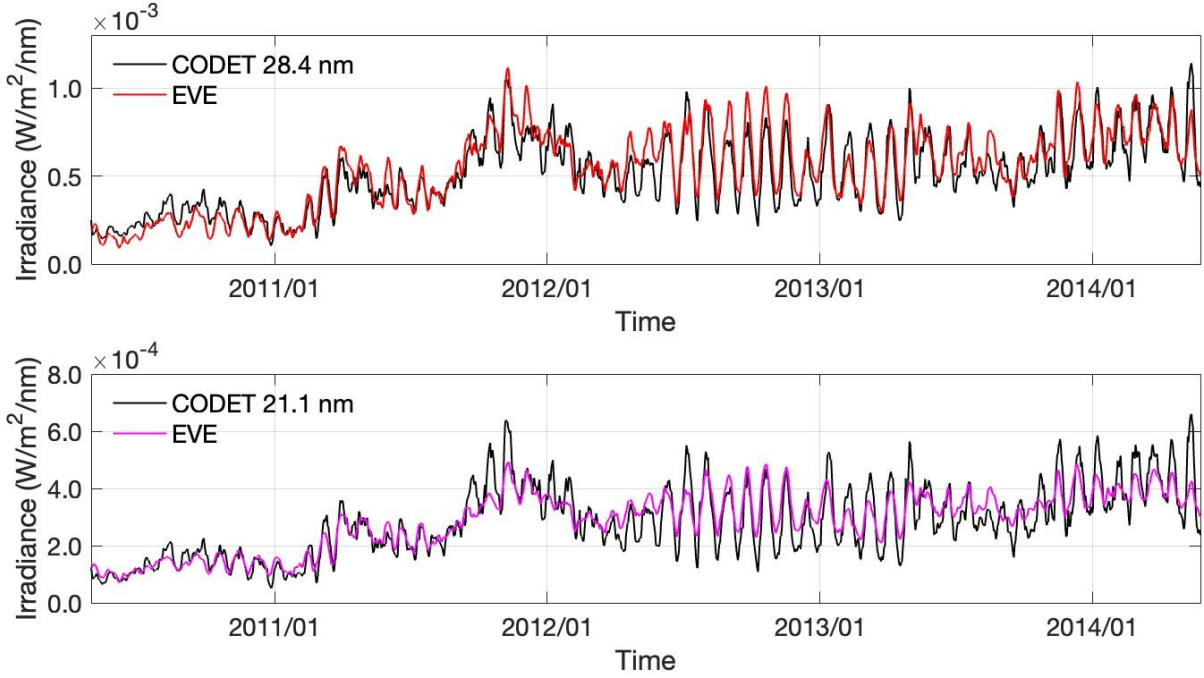

**Figure 2.** Solar Spectral Irradiance (SSI) using CODET model version 1.1 (black line) and Solar Spectral Irradiance from EVE/SDO at 28.4 nm (red line) and 21.1 nm (magenta line) from April 30, 2010 to May 26, 2014.

**Table 1.** CODET model version 1.1 parameters: scaling law exponents for density $\gamma$ and temperature $\alpha$, background density $N_o$ [$cm^{-3}$] and temperature $T_o$ [K], $B_s$ [G] is a constant value of the magnetic field, Goodness-of-fit $\chi^2$, and specifications about the optimization algorithm: population size, and generation.

| Parameter | Value | Units |
|---|---|---|
| $\gamma$ | 1.3077 | ... |
| $\alpha$ | −0.2781 | ... |
| $N_o$ | $8.2819 \times 10^8$ | $cm^{-3}$ |
| $T_o$ | $1.8558 \times 10^6$ | K |
| $B_s$ | 7.9966 | G |
| $\chi^2$ | $1.4502 \times 10^{-2}$ | ... |
| Population size | 20 | ... |
| Generation | 50 | |

The power-law exponents for density $\gamma$ and temperature $\alpha$, background density $N_o$ [$cm^{-3}$] and temperature $T_o$ [K], and the constant value of magnetic field $B_s$ [G] that drive the amount of magnetic flux in each pixel (more details Appendix A).

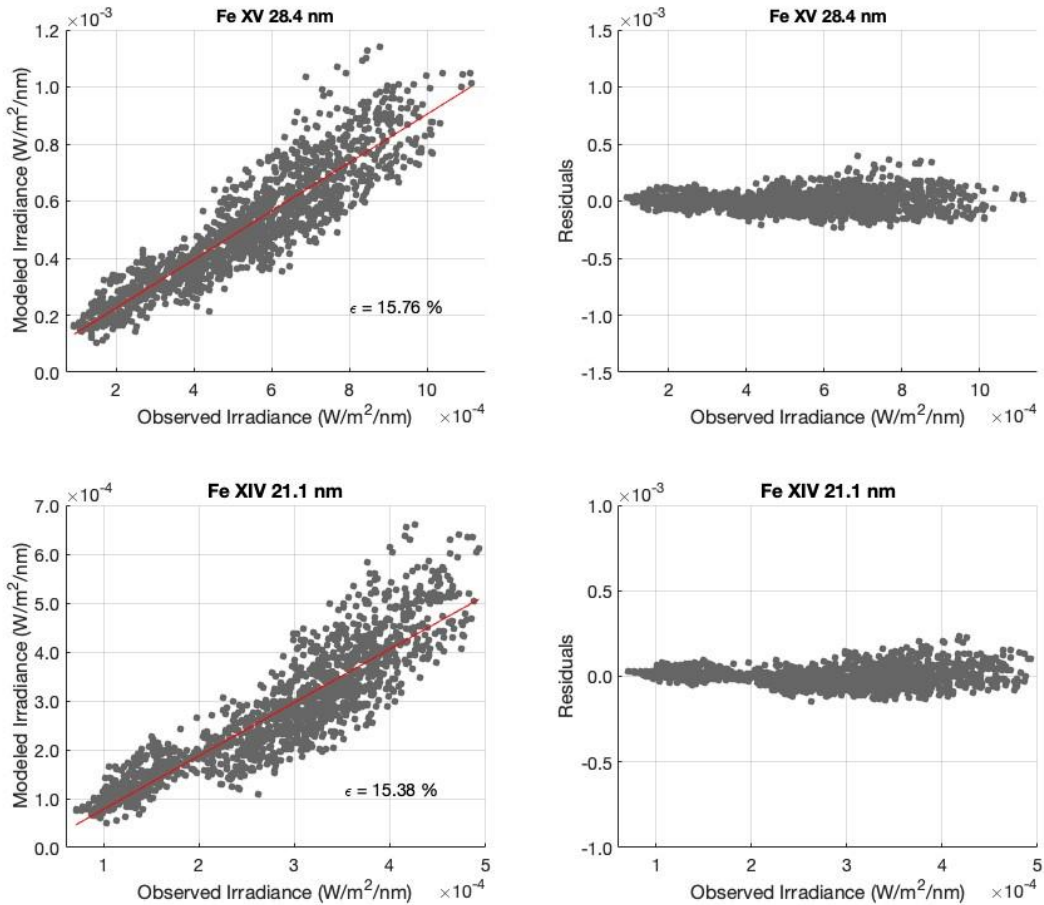

**Figure 3.** Solar Spectral Irradiance observed by SDO/EVE and modeled using the CODET model version 1.1 at 28.4 nm and 21.1 nm wavelengths from April 30, 2010 to May 26, 2014 scatter plots with the Mean Absolute Percentage Error (MAPE) $\epsilon$ (left column) and residuals (right column).

Figure 3 compares observed irradiance from SDO/EVE and modeled irradiance from the CODET model version 1.1 at 28.4 nm and 21.1 nm wavelengths (left column). A linear fit (red line) was overplotted in each scatter plot with $R^2$ = 0.849 and $R^2$ = 0.796 for Fe XV 28.4 *nm* and Fe XIV 21.1 *nm*, respectively. The residuals between modeled and observed data are shown in the right column. The norm of residuals corresponds to 0.0033 for Fe XV 28.4 *nm* and 0.0021 for Fe XIV 21.1 nm. These results show that CODET model version 1.1 describes well both wavelengths. The $\chi^2$ value was obtained using equation A6 in the Appendix, $\chi^2$ values for 28.4 nm and 21.1 nm correspond to 0.0250 and 0.0153, respectively. It is important to highlight that the model's performance adequately describes the Solar Spectral Irradiance when observational data like SDO/EVE MEGS-A is used to obtain the goodness-of-fit in the model (Figure 1). In addition, it is reflected through the Mean Absolute Percentage Error (MAPE) $\epsilon$ obtained with the *mape* function in Matlab[4], in both cases $\epsilon \sim 15\%$, showing how accurate a model's predictions are. Also, the relative irradiance error $\left(\frac{I_{model}-I_{obs}}{I_{obs}}\right) \times 100\%$ was obtained from CODET model v1.1 and observed SSI from SDO/EVE at 28.4 nm and 21.1 nm from Figure 2. The error analysis is displayed in Figure 4 between modeled and observed SSI (left column) and their corresponding histograms (right column); the dotted red line shows $2\sigma^2$ values, indicating that 95% of the SSI values within that range describe the best model performance, and values outside that limit show intervals where the CODET model v1.1 was not able to fit the observational data, e.g., 28.4 nm shows a poor fitting from May 17, 2010, to October 27, 2010, with $2\sigma > 41.35\ \%$, but in general the CODET model v1.1 correctly describes SSI in 28.4 nm from November 21, 2010 to May 26, 2014, where those values lie in the interval $\pm 2\sigma$. Also, the model describes competently the SSI in 21.1 nm through the relative error values; almost all of them are between $2\sigma \pm 38.00\ \%$, except some days that exceed that value.

---

[4] https://www.mathworks.com/help/matlab/ref/mape.html

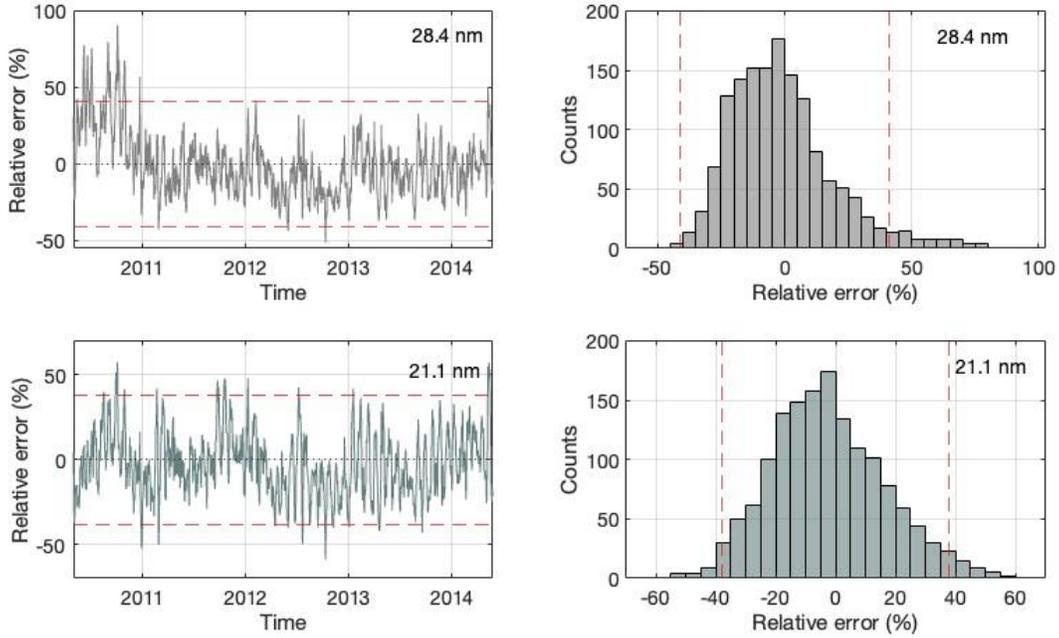

**Figure 4.** Relative error (%) between CODET model v1.1 and observed SSI from SDO/EVE at 28.4 nm (top) and 21.1 nm (bottom) for the 2010 to 2014 period, and their respective histograms, dotted red lines show $2\sigma^2_{28.4nm} = \pm\ 41.35\ \%$ and $2\sigma^2_{21.1nm} = \pm\ 38.00\ \%$ values.

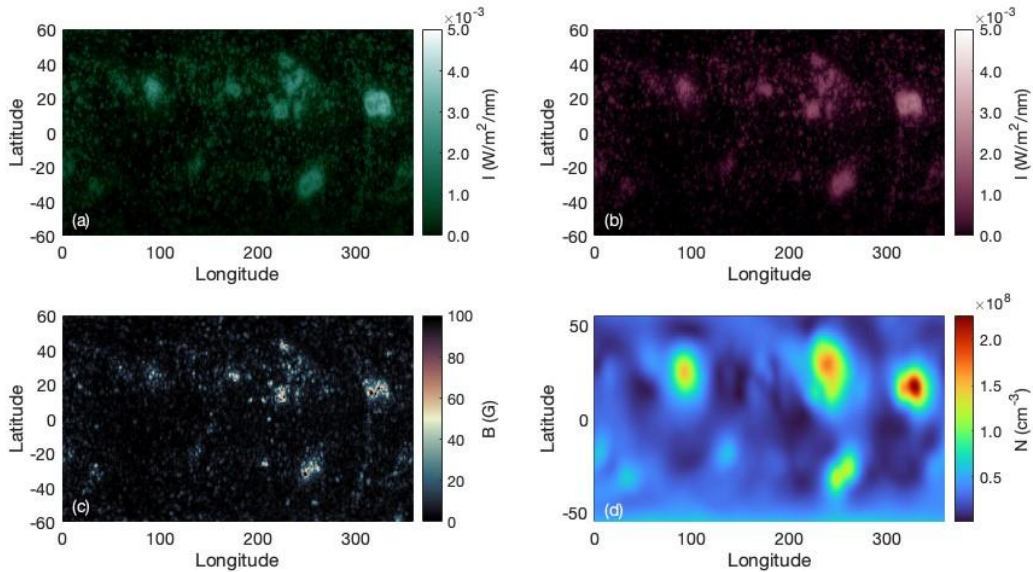

**Figure 5.** Full-disc intensity maps from CODET model version 1.1 at 28.4 nm (a) and 21.1 nm (b), photospheric magnetic field map (c), and density map at 1.16 $R_\odot$ (d) on May 2, 2010.

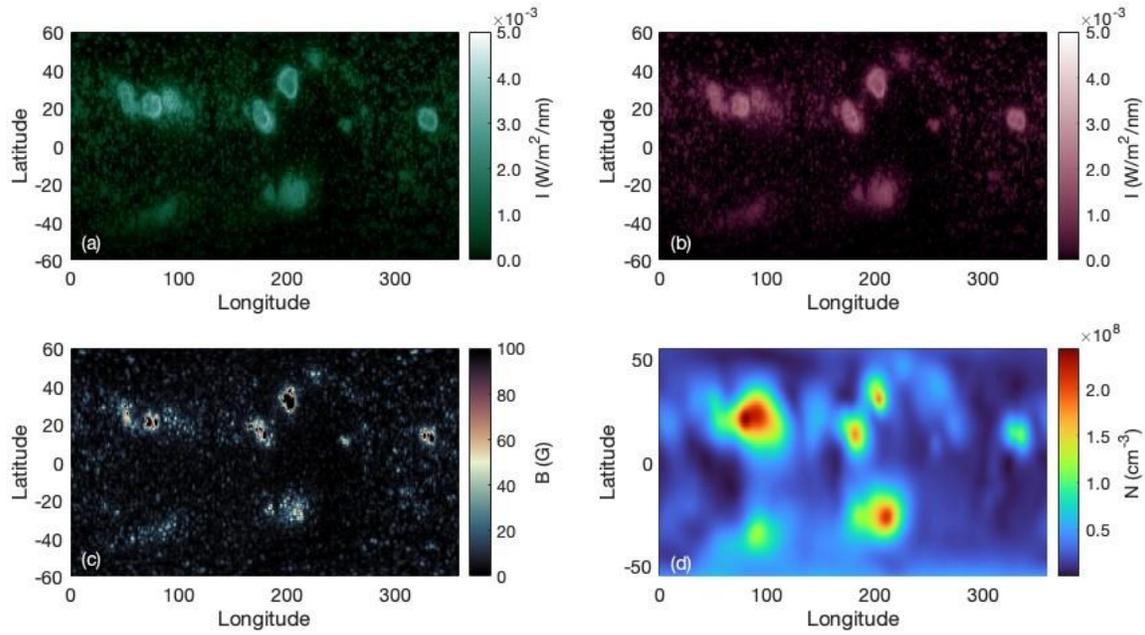

**Figure 6.** Full-disc intensity maps from CODET model version 1.1. at 28.4 nm (a) and 21.1 nm (b), photospheric magnetic field map (c), and density map at 1.16 R$_\odot$ (d), on December 14, 2010.

Figures 5 and 6 show two examples of full-disc intensity maps (new product) in Fe XV 28.4 nm and Fe XIV 21.1 nm from CODET model v1.1, the photospheric magnetic field maps (Schrijver 2001) using SDO/HMI data, and density maps at a height of 1.16 R$_\odot$ on May 2, 2010 and December 14, 2010. Those days were selected because they represent a goodness-of-fit between the observational data and CODET model v1.1 (Figure 2). These intensity maps allow a direct comparison with observational data from different missions, e.g., SDO/AIA and STEREO/EUVI. Also, the density maps provide information about the plasma variation at different coronal heights on features present in the solar atmosphere like quiet sun, active regions, coronal holes etc.

### 3.1. *Comparison between CODET model versions 1.0 and 1.1*

The comparison between both model versions is possible uniquely using 21.1 nm, the wavelength that both versions have in common. Figure 7 shows CODET model version 1.0 at 21.1 nm from April 30, 2010, to May 26, 2014, obtained using the model parameters from Rodríguez Gómez et al. (2018) and plotted together with TIMED/SEE level 3 version 12 at 21.5 nm, scatter plots, and residuals. In general, the CODET model version 1.0 mean intensity value corresponds to $1.1530 \times 10^{-4}$ *W/m$^2$/nm* while the CODET model version 1.1 mean value is $2.8306 \times 10^{-4}$ *W/m$^2$/nm* during the period of interest. The CODET model version 1.0 shows a mean SSI value lower than ∼ 41% compared with the updated version of the CODET model version 1.1. The main reason behind that is the data used to obtain the Goodness-of-fit in version 1.0 was TIMED/SEE; as was mentioned before, this dataset is not real observational data. This is also evident in the mean intensity ratio between EVE and TIMED, which was presented previously in section 2.

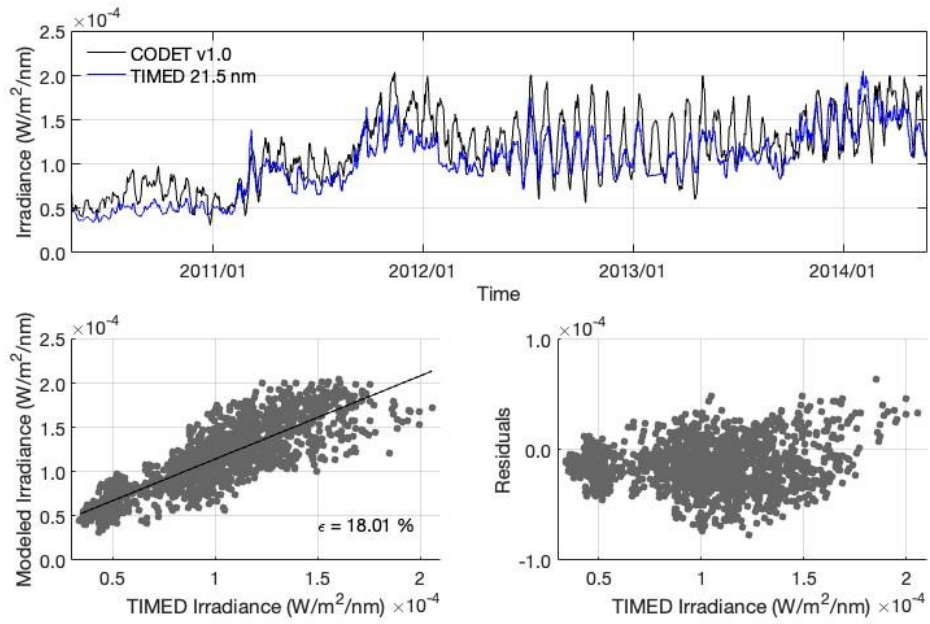

**Figure 7.** SSI from CODET model version 1.0 and TIMED/SEE from April 30, 2010, to May 26, 2014 (top), scatter plots with the mean absolute percentage error ($\epsilon$) and residuals (bottom).

The scatter plot shows consistency between the modeled intensity from CODET v1.0 and TIMED/SEE with an error of 18.01%, which is 2.63% more than the error obtained when CODET model version 1.1 and SDO/EVE were compared. Also, residuals show a big scatter in the model's oldest version (Figure 7). It highlights the significant improvement of the new model version, together with the new wavelength (28.4 nm) included, full-disc images and predictions closer than the observed data. However, both model versions adequately describe SSI using observational (SDO/EVE) and observational-model (TIMED/SEE) data.

### 3.2. *Solar Spectral Irradiance prediction*

Due to the relationship between magnetic field and emission in EUV, it is possible to obtain SSI estimations from July 01, 1996, to October 28, 2024 (time of writing), where SOHO/MDI and SDO/HMI photospheric magnetic field data are available. This interval covers solar cycles 23 and 24 and the ascending phase of solar cycle 25. Figure 8 shows SSI predictions from the CODET model v1.1 at 28.4 nm (red line) and 21.1 nm. EUVS daily averages from GOES-16 data[5] at 28.4 nm and SDO/AIA full disc mean daily intensity at 21.1 nm were over-plotted from June 05, 2017, to October 27, 2024, and from May 20, 2010, to November 18, 2022, respectively. Also, some outliers (peaks) appearing on the SSI prediction on August 16, 2016, January 4, 2019, and May 1, 2023, correspond to artifacts on the magnetic flux transport maps from Schrijver (2001) [6].

---

[5] https://www.ngdc.noaa.gov/stp/satellite/goes-r.html

[6] available at http://www.lmsal.com/solarsoft/archive/ssw/pfss_links v2/

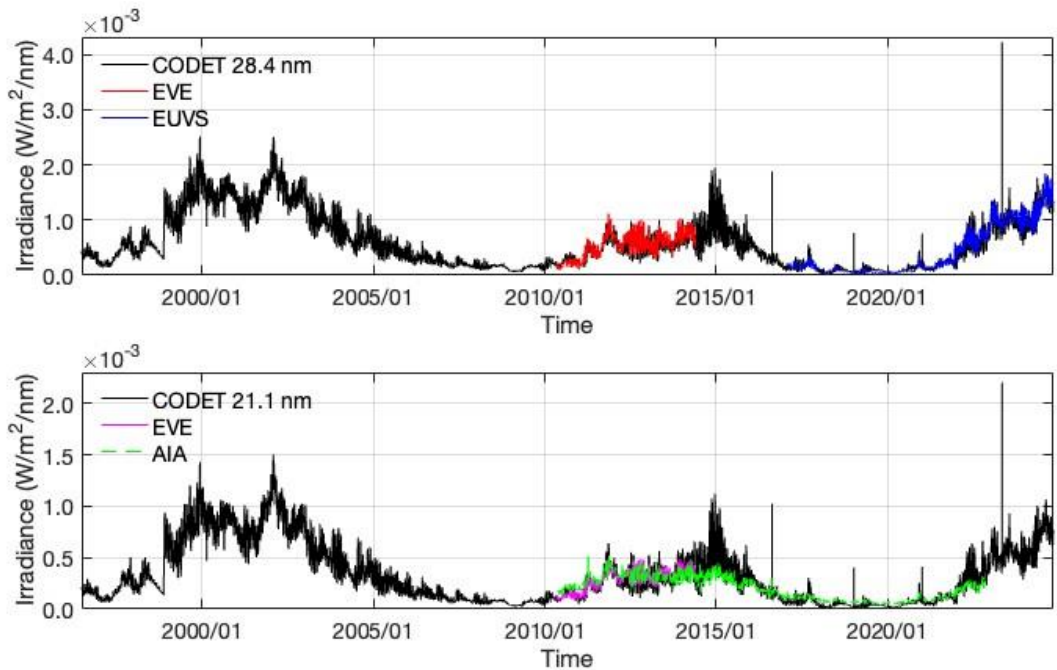

**Figure 8.** Solar Spectral Irradiance (SSI) prediction using COronal DEnsity and Temperature model v1.1 from July 01, 1996, to October 28, 2024. Top: Solar Spectral Irradiance from CODET version 1.1 at 28.4 nm (black line), EVE/SDO (red line), and EUVS/GOES (blue line). Bottom: SSI from CODET version 1.1 at 21.1 nm (black line) and EVE/SDO (magenta line) and AIA daily mean values (green line) were overplot.

Scatter plots and residuals of CODET v1.1 predictions and observed SSI from GOES/EUVS at 28.4 nm from June 5, 2017, to October 27, 2024, and SDO/AIA daily mean values at 21.1 nm from May 20, 2010, to November 18, 2022, and the mean absolute percentage error are shown in Figure 9. Although the primary goal of using daily mean SDO/AIA values was to make a qualitative comparison between observed daily mean SSI and the model prediction performance, an error analysis was performed. The Mean Absolute Percentage Error correspond to $\epsilon \sim 26\%$ and $42\%$ for 28.4 nm and 21.1 nm, respectively. Additionally, the $\chi^2$ value was obtained to compare observational data with CODET model version 1.1 predictions. A value of $\chi^2 = 0.0498$ and $\chi^2 = 0.364$ were obtained for 28.4 nm and 21.1 nm, respectively. High error ($\epsilon$) and $\chi^2$ values were obtained when CODET model version 1.1, and SDO/AIA were compared. It is due to some factors, e.g., to compare SSI from CODET version 1.1 and AIA was required to divide AIA full disc daily mean values by an empirical value of $7 \times 10^5$ to scale with the SSI at 21.1 nm. The main reason to use this empirical value is to compare two different kinds of instruments and obtain a reliable value that allows to compare SSI values in different units, AIA in [*DN*] and modeled irradiance in [*W/m²/nm*]. Also, it is important to highlight that instrumental characteristics like the bandpass and the response functions of each instrument can affect the direct comparisons for AIA and GOES/EUVS daily values with the model outputs.

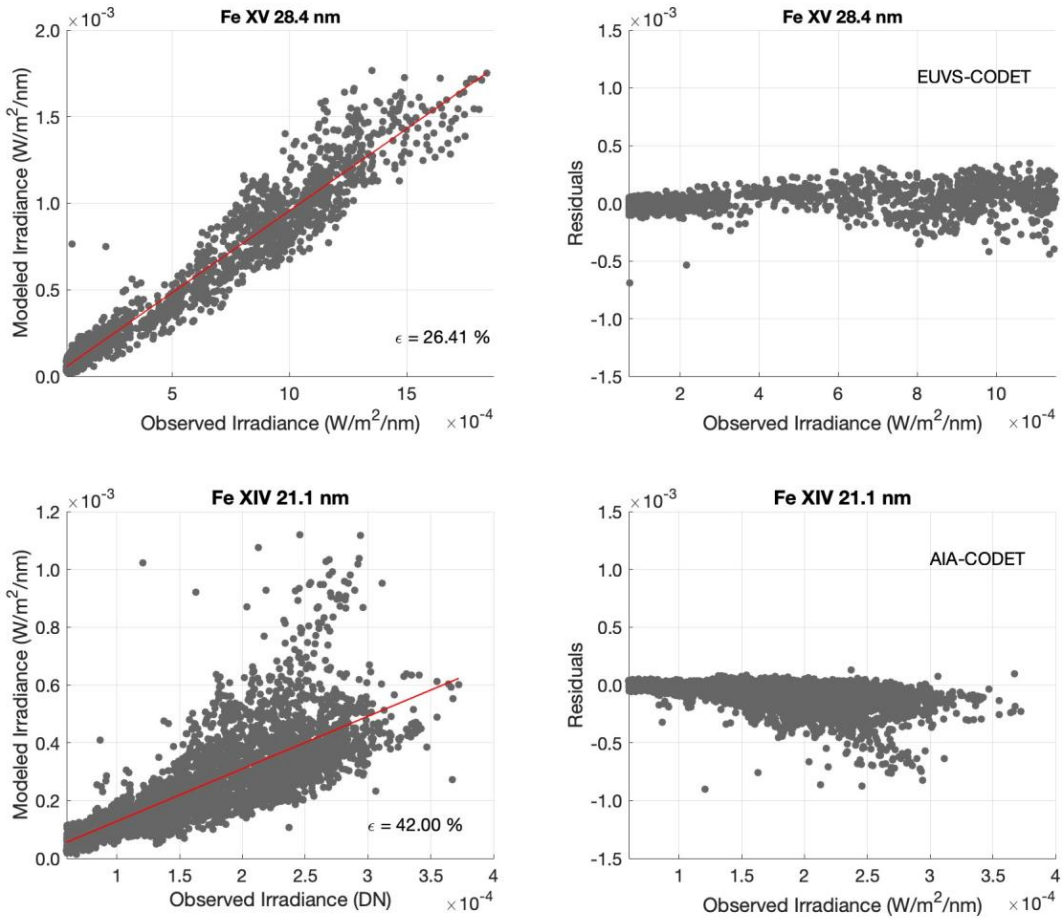

**Figure 9.** Solar Spectral Irradiance observed by GOES/EUVS from June 5, 2017 to October 27, 202, SDO/AIA from from May 20, 2010 to November 18, 2022 and the CODET model version 1.1 predictions at 28.4 nm and 21.1 nm wavelengths scatter plots with the Mean Absolute Percentage Error ($\epsilon$) (left column) and residuals (right column).

## 4. DISCUSSION AND CONCLUDING REMARKS

The COronal DEnsity and Temperature (CODET) model version 1.1 describes the emission of the hottest lines in the solar corona, such as 28.4 nm and 21.1 nm. It allows for obtaining SSI estimations in EUV wavelengths where the photospheric magnetic field is available, helping to cover observational gaps in the SSI time series. The main results of this study can be summarized as follows.

- Modeled Solar Spectral Irradiance was obtained by the CODET model version 1.1 (Figure 2) covering the interval from April 30, 2010, to May 26, 2014, EVE/SDO- MEGS-A detector observation time interval in 28.4 nm compared to the 21.1 nm wavelengths. It confirms that it is possible to retrieve SSI from the hottest lines in solar corona, e.g., Fe XIV and Fe XV lines, using the relationship between magnetic field, density, temperature, and emission described thought the CODET model. By examining the residuals, scatter plots and error analysis on Figures 3 and 4 at 28.4 nm and 21.1 nm are well described in this model version. It can be seen through the mean $\chi^2$ value ~ 0.0202 and the error analysis using the mean absolute percentage error and the relative error, which show errors less than 20% in both wavelengths, highlighting the model's good performance. Using the relative error (Figure 4) and specifically the ± $2\sigma^2$ limits, e.g., $2\sigma^2_{28.4nm} = \pm 41.35\%$ and $2\sigma^2_{21.1nm} = \pm 38.00\%$ is possible to evaluate where the CODET model v 1.1 works well and in which days or intervals it does not occur. In some cases, the model overestimate or underestimate SSI exceeding $2\sigma^2$ values in both wavelengths, e.g., October 16, 2011, November 6, 2011, May 4, 2013,

November 19, 2013, and May, 12 2014, show big and complex Active Regions covering the solar disc (Figures [2] and [4]). These wavelengths allow to observe upper coronal heights and hottest loops related to these complex ARs, and as the model describes their magnetic structure using a PFSS magnetic field extrapolation, maybe it is not enough to describe them. However, the CODET model v 1.1 describes the SSI adequately from April 30, 2010, to May 26, 2014, during the ascending phase - maximum solar cycle 24. This means that, in general terms, this model describes emission in EUV during times where the magnetic activity increases until reaching the solar maximum, despite some cases, as mentioned before.

- Even though the model parameters used to obtain density and temperature in scaling laws and subsequently used as input of the emission model described through the contribution function ($G(\lambda,T)$) obtained from CODET model v 1.1 (Table [1]) differs from the previous version of the model. The main reasons are the inclusion of a new wavelength of 28.4 nm and using observational data (SDO/EVE) to obtain the goodness-of-fit in this updated version. Nevertheless, the model parameter values correspond to $\gamma > 0$, $\alpha < 0$, and $B_s < 10\ G$, as was obtained in the oldest model version (CODET v 1.0).

- The meaning and validation of the density and temperature estimations were investigated through the plasma $\beta$ (Rodríguez Gómez et al. [2019]), showing how the CODET model parameters v1.0 represent well the dynamics of the solar corona during solar cycles 23 and 24. The updated version of the CODET model version 1.1 (Figures [5] and [6]) show mean density values ∼ $10^8\ cm^{-3}$ in agreement with the values reported by Rodríguez-Gómez et al. ([2024]) found through observational constraints from Hinode/EIS and SDO/AIA in the quiet sun corona. In addition, it is planning to compare density profiles from CODET model version 1.1 with density profiles obtained from observational data, e.g., Metis/SoLO, using modeled SSI from CODET model v1.1 in wavelengths available in EVE/SDO MEGS-B detector and/or EUVS/GOES more contemporaneous with SoLO mission. As is well known, electron density and temperature are important for describing the solar corona and its physical processes. Density changes with height are essential for modeling the acceleration of the fast solar wind and for providing some descriptions related to mechanical forces and energy in the solar corona (Mercier & Chambe [2015]; Lallement et al. [1986]; Fludra et al. [1999]). Thus, density maps and profiles obtained from CODET model version 1.1 between 1 R$_\odot$ to 2.5 R$_\odot$ cover the lower and inner corona, which becomes very valuable in that context.

- The CODET model version 1.1 provides full-disc maps as a new product (e.g., Figures [5] and [6]) in addition to the Solar Spectral Irradiance daily time series in 28.4 nm and 21.1 nm wavelengths. Those maps can be provided daily when the photospheric magnetic field is available. The importance of these maps relapses in the option to provide maps where no observational data is available, e.g., when STEREO Ahead EUVI images at 28.4 nm are not available (where STEREO is observing the Sun on the far-side).

- The comparison between the CODET model version 1.0 and the updated version 1.1 at 21.1 nm from April 30, 2010, to May 26, 2014, allow to observe how the time series used to obtain the goodness-of-fit affect the model outputs. Thus, in the updated version of this model, when the real observational data (SDO/EVE) was used, it allowed to provide reliable estimations of SSI at 28.4 nm and 21.1 nm. However, both model versions provide a good description of SSI variability covering the last solar cycles. Hence, CODET model versions 1.0 and 1.1 adequately describe SSI using observational and observational-model data, e.g., SDO/EVE and TIMED/SEE in wavelengths of 19.5 nm, 21.1 nm, and 28.4 nm. This implies that the physical descriptions behind the model, like the relationship between density, temperature, magnetic field, and emission are suitable for reliable estimations, even where no SSI observational data is available (Figure [8]).

- The updated CODET model version allows to provide SSI predictions in 28.4 nm and 21.1 nm where SOHO/MDI and SDO/HMI photospheric magnetic field data are available. The model predictions presented in this work cover the period between July 01, 1996, and October 28, 2024. These long and continuous time series are important to analyze the long-term variability of SSI in EUV. It is important to highlight that the magnetic field and the flux transport models can affect these estimations. For example, they can cause some outlier or extreme SSI values (Figure [8]), where some artifacts appear on the flux transport magnetic flux maps. These predictions were compared with GOES/EUVS and SDO/AIA data. The comparison between SDO/AIA 21.1 nm and CODET model prediction shows an error of ∼ 42% related to the comparison from the daily average intensity in DN units and SSI from the model in $W/m^2/nm$, the use of an empirical value to scale both values and the instrumental characteristics. However, the comparison with GOES/EUVS at 28.4 nm provides an error of ∼ 26%, highlighting the good performance of the CODET model v1.1.

- This model shows an incredible potential. In the future, more wavelengths are expected to be included in the model scope to obtain the SSI, which will provide a good description of the SSI variability in long time scales from days to solar cycles. Additionally, Solar Spectral Irradiance from the CODET model can be used as input in Ionosphere-Thermosphere-Mesosphere (ITM) models to investigate the impact of Solar Spectral Irradiance in EUV wavelengths in the different Earth's atmospheric layers.

This work was supported by NASA Living With a Star (LWS) Program, Focused Science Topic: "Beyond F10.7: Quantifying Solar EUV Flux and its Impact on the Ionosphere – Thermosphere – Mesosphere System" No. 80NSSC23K0900.

## APPENDIX

### A. MODEL PARAMETERS DESCRIPTION

The CODET model parameters description was present previously by Rodríguez Gómez et al. (2018). A brief description is presented below.

The density and temperature profiles are described as a function of the magnetic field using scaling laws. The magnetic field B [G] comes from the PFSS extrapolation

$$B = \sqrt{B_r^2 + B_\phi^2 + B_\theta^2} \tag{A1}$$

Thus, the density and temperature distribution can be defined as

$$N(B) = N_o \left(\frac{B}{B_s}\right)^\gamma \quad [cm^{-3}] \tag{A2}$$

The function $B_f(R) = b_{fo} \times e^{\left(\frac{R}{\tau_{bf}}\right)^2}$ where $b_{fo}$ and $\tau_{bf}$ are constant values and $R$ is the height through the solar corona. Rodríguez Gómez et al. (2018) found $b_{fo} = 20\,G$, $\tau_{bf} = 1.2 R_\odot$ and R from 1 $R_\odot$ to 2.5 $R_\odot$. $B_f(R)$ was defined to describe two different temperature regimes related to regions with strong and weak magnetic field, respectively. Thus, the temperature can be defined as if $B < B_f(R)$

$$T(B) = T_o \tag{A3}$$

if $B > B_f(R)$

$$T(B) = T_o \left(\frac{B}{B_s}\right)^\alpha \quad [cm^{-3}] \tag{A4}$$

The parameters $\gamma$ and $\alpha$ are power-law indices, $\left(\frac{B}{B_s}\right)$ is the factor related to the amount of flux in each pixel, $B_s$ [G] is a constant value of the magnetic field, and $N_o$ [cm$^{-3}$] and $T_o$ [K] are the background density and temperature, respectively. The parameters $\gamma$, $\alpha$, $B_s$, $N_o$ and $T_o$ are the free parameters in the model. The density and temperature profiles are used as input in the emission measure formalism. Assuming that the emission lines are optically thin, the specific intensity can be described by

$$I_o(\lambda) = \iint R(\lambda)\, G(\lambda, T)\, d\lambda\, N^2 ds \tag{A5}$$

where $G(\lambda,T)$ is the contribution function from the CHIANTI atomic database[7] 10.0.2, $d\lambda$ is the differential element in wavelength, $ds$ is the differential distance along the line of sight and $R(\lambda)$ is the instrumental response and in this description is consider equal to one.

The Goodness-of-fit ($\chi^2$) used to compare SSI from CODET model ($I_{model}$) with the observational SSI time series ($I_{obs}$) was defined as

$$\chi^2 = \frac{(I_{model} - I_{obs})^2}{|I_{obs}|} \tag{A6}$$

---

[7] https://www.chiantidatabase.org/